\documentclass[twocolumn,showpacs,preprintnumbers,amsmath,amssymb]{revtex4}
\usepackage{graphicx}% Include figure files
\usepackage{dcolumn}% Align table columns on decimal point
\usepackage{bm}% bold math

\begin{document}

\preprint{hep-th/0307276}

\title{Solvable Models of Domain Walls 
%Coupled with 
in ${\cal N}=1$ 
Supergravity}% Force line breaks with \\

\author{Minoru Eto}
% \altaffiliation[Also at ]{Physics Department, XYZ University.}%Lines break automatically or can be forced with \\
 \email{meto@th.phys.titech.ac.jp}
\author{Norisuke Sakai}%
 \email{nsakai@th.phys.titech.ac.jp}
\affiliation{%
Department of Physics, Tokyo Institute of Technology,\\
%Authors' institution and/or address\\
Oh-okayama, Meguro, Tokyo 152-8551, Japan
%This line break forced with \textbackslash\textbackslash
}%

%\author{Charlie Author}
% \homepage{http://www.Second.institution.edu/~Charlie.Author}
%\affiliation{
%Second institution and/or address\\
%This line break forced% with \\
%}%

\date{\today}% It is always \today, today,
             %  but any date may be explicitly specified

\begin{abstract}
 A class of exactly solvable models of domain walls 
are worked out in $D=4$ ${\cal N}=1$ supergravity. 
 We develop a method to embed globally 
 supersymmetric theories with exact BPS domain wall solutions 
into supergravity, by introducing a gravitationally deformed 
superpotential. 
The gravitational deformation is natural in the spirit of 
maintaining the K\"ahler invariance. 
The solutions of the warp factor and the Killing spinor 
are  also obtained.  
We find that three distinct behaviors of warp factors arise 
depending on the value of a constant term in the superpotential 
: 
exponentially decreasing in both sides of the wall, 
flat in one side and decreasing in the other, and 
increasing in one side and decreasing in the other. 
Only the first possibility gives the localized massless 
graviton zero mode. 
Models with multi-walls and 
models with runaway vacua 
are also discussed. 
\end{abstract}

% PACS, the Physics and Astronomy
                             % Classification Scheme.
%\keywords{Suggested keywords}%Use showkeys class option if keyword
                              %display desired
\maketitle

\section{Introduction}

The domain wall has been an interesting subject in many 
areas of physics, such as particle 
physics, cosmology and the condensed matter physics. 
In particle physics,  the recently developed brane-world 
models \cite{LED}--\cite{RS2} give a new motivation to 
study domain walls. 
On the other hand, supersymmetry (SUSY) has been most 
useful to build 
unified models beyond the standard model \cite{DGSW} 
and helps to construct domain walls. 
Configurations preserving part of SUSY are called 
Bogomolo'nyi-Prasad-Sommerfield (BPS) 
configurations \cite{BPS, WittenOlive}, which 
automatically give 
solutions of equations of motion 
with minimum energy for the given boundary conditions. 

Exact solutions are useful to understand solitons such as 
domain walls. 
There have been a number of works to obtain exact solutions 
in models with global SUSY. 
In ${\cal N}=1$ SUSY models, 
%with chiral scalar fields (Wess-Zumino models), 
exact solutions for a single wall are abundantly available 
\cite{SingleWall}, 
and also for two-wall solutions with a moduli parameter 
in a model with two chiral scalar fields 
\cite{Shifman:1997wg, GK}. 
Interesting dynamics of multi-walls in this model 
are also discussed 
\cite{SakaiSugisaka, AGMT}. 
Even in ${\cal N}=2$ SUSY models (with eight SUSY), exact single wall 
solution \cite{GPT, ANNS} 
as well as exact multi-wall solutions have been constructed 
\cite{GPT}. 
Exact solutions of domain-wall junctions have also been found 
for ${\cal N}=1$ models \cite{OINS, NNS2} 
and for ${\cal N}=2$ models \cite{KakimotoSakai}. 

On the other hand, it has been difficult to obtain 
exact solutions in 
supergravity (SUGRA), because of highly non-linear nature of gravity. 
Many attempts revealed useful qualitative features of 
domain walls in SUGRA theories in four- 
and five-dimensions \cite{CQR}--\cite{Eto:2003ut}. 
Recently, exact domain wall solutions in SUGRA with a smooth limit of 
weak gravity have been 
found in several models : a periodic model in ${\cal N}=1$ SUGRA in 
four-dimensions \cite{Eto:2002ns}, and $T^*{\bf C}P^n$ models 
in ${\cal N}=2$ SUGRA in five-dimensions \cite{Arai:2002ph, Eto:2003ut}. 
Interestingly, the scalar field 
configurations in all these SUGRA solutions are found to be 
exactly the same as the known solution in global SUSY models 
(the limit of the vanishing gravitational coupling).

The purpose of this paper is to propose a general method to find 
exact solutions of domain wall in SUGRA models. 
Inspired by the exact solutions in the above models 
\cite{Eto:2002ns, Arai:2002ph, Eto:2003ut}, we find conditions 
to preserve the scalar field 
configurations of the exact solutions in global SUSY models 
when they are embedded into SUGRA. 
Namely, we require that the scalar field configurations be 
unchanged when we find the distortion of the spacetime 
together with the backreaction by solving 
the nonlinear field equations of SUGRA with matter. 
Thus we obtain the 
necessary gravitational deformations to the superpotential. 

In general the SUSY vacua in global SUSY theories change 
when the theory is coupled to SUGRA due to gravitational effects. 
This is one of the reasons which prevent 
us from obtaining the exact domain wall solutions. 
Therefore our main strategy is that 
we require gravitational deformations of the superpotential 
when it is embedded into SUGRA so that the SUSY vacua remain 
unchanged. 
As a result, we find that the modified superpotential gives us 
precisely the same equation for the scalar field as the one in the 
global SUSY theory. 
Therefore we obtain the solution for the scalar 
field configuration which is identical to the global SUSY theory. 
Once we obtain the 
exact domain wall solutions, we can also find the distortion of the
spacetime by solving the 
%other independent 
equations for the spacetime 
metric. 
We can also obtain the Killing spinor corresponding to the
conserved SUSY.

In Sec.\ref{sc:SUGRA-SUSY} we briefly 
review domain walls in SUGRA and discuss the
deformation of the superpotential which maintains the wall 
configuration identical to the global SUSY case. 
%against the gravitational effect. 
A number of interesting models with exact BPS solutions are 
described in Sec.\ref{sc:solvable-examples}. 
The zero modes and the warp factor are discussed in 
Sec.\ref{sc:zero-mode}. 
Sec.\ref{sc:conclusion} is devoted to concluding remarks.

\section{Domain walls in ${\cal N}=1$ SUGRA}
\label{sc:SUGRA-SUSY}
\subsection{Notations}
We consider $n$ chiral multiplets $(\phi^i,\chi_\alpha^i)$ 
coupled to the gravity multiplet 
$(e_m{^{\underline{a}}},\psi_m{^\alpha})$ 
with the K\"ahler potential $K(\phi,\phi^*)$ and the superpotential
$P_{\rm lc}(\phi)$ in four dimensions. 
In order to distinguish the superpotential 
in the SUGRA theories from that in the global 
SUSY theories,
we denote the superpotential in the local
SUSY (SUGRA) as $P_{\rm lc}$ and 
the superpotential in the global SUSY theories as $P_{\rm gl}$. 
The local Lorentz vector indices are denoted by letters with the
underline as $\underline{a}$, and the vector indices transforming under
general coordinate transformations are denoted by Latin letters as
$m,n=0,1,2,3$. The left(right)-handed spinor indices are denoted by
undotted (dotted) Greek letters as $\alpha(\dot\alpha)$.
We follow Ref.\cite{Wess:cp} about other notations for the spinor algebra.
The bosonic part of the Lagrangian of 
the ${\cal N}=1$ SUGRA coupled with the chiral 
multiplets is given by \cite{Wess:cp} : 
\begin{eqnarray}
e^{-1}{\cal L}
&=& - \frac{1}{2\kappa^2} R
  - K_{ij*}g^{mn}\partial_m\phi^i\partial_n\phi^{j*}
  - V_{\rm lc},\label{main_lagrangian}
\end{eqnarray}
where the gravitational coupling $\kappa$ is the inverse of the 
four-dimensional Planck mass $M_{\rm Pl}$, $g_{mn}$ is the metric of 
the spacetime, $e$ is the determinant of the vierbein 
$e_m{^{\underline{a}}}$. 
The scalar potential in the SUGRA theory is denoted by $V_{\rm lc}$ 
and is given in terms of the covariant derivative $D_iP_{\rm lc}$ 
in the target space of the superpotential $P_{\rm lc}$ in 
SUGRA 
\begin{eqnarray}
V_{\rm lc}
&=& {\rm e}^{\kappa^2K}
    \left(K^{ij*}D_iP_{\rm lc} D_{j*}P^*_{\rm lc} 
- 3\kappa^2 |P_{\rm lc}|^2\right),
\label{scalar_pot}
\end{eqnarray}
\begin{eqnarray}
D_i P_{\rm lc}(\phi) 
= \partial_iP_{\rm lc}(\phi) + \kappa^2P_{\rm lc}(\phi)\partial_iK(\phi,\phi^*).
\label{CD_SUGRA}
\end{eqnarray}
%In the above Lagrangian (\ref{main_lagrangian}) we turn off all 
%the fermionic fields as a tree level configuration. 

The SUGRA Lagrangian %(\ref{main_lagrangian}) 
is invariant under the SUGRA 
transformation. 
SUSY vacua and BPS solutions can be obtained by examining 
SUGRA transformations. 
Since we are interested in classical solutions, 
fermionic fields should vanish. 
Therefore we need to consider 
SUGRA transformations of fermions which 
read 
%\footnote{We have little interest in the transformation law for the
%bosonic fields since it is trivial for our solutions.}:
\begin{eqnarray}
\begin{array}{l}
\delta_\zeta\psi_m 
= 2\kappa^{-1}{\cal D}_m\zeta 
    + i\kappa {\rm e}^{\frac{\kappa^2}{2}K}P_{\rm lc}\sigma_m\bar\zeta,\\
\delta_\zeta \chi^i
= i\sqrt{2}\sigma^m\bar\zeta\partial_m\phi^i
- \sqrt{2} {\rm e}^{\frac{\kappa^2}{2}K} 
K^{ij*}D_{j*}P_{\rm lc}^*\zeta,
\end{array}\label{SUSY_transf}
\end{eqnarray}
where we drop terms which include the fermionic fields and 
$\zeta$ is a local SUSY transformation parameter. 
The spacetime covariant derivative is given by 
\begin{eqnarray*}
{\cal D}_m\zeta 
= \partial_m\zeta + \zeta\omega_m
  + \frac{i\kappa^2}{2}\sum_i
{\rm Im}\left[\partial_iK\partial_m\phi^i\right]\zeta,
\end{eqnarray*}
where $\omega_m$ denotes the spin connection.

It is well known that a stable solitonic state can be obtained, 
if part of SUSY (or SUGRA) is 
preserved and the BPS energy bound is saturated 
\cite{Cvetic:1992bf, SingleWall}. 
 The  domain wall solutions interpolating two isolated SUSY vacua 
typically preserve two out of four SUSY (or SUGRA) and are called 
$\frac{1}{2}$ BPS states. 
Let us parametrize the conserved directions of SUGRA transformations as 
\begin{eqnarray}
\zeta(y) = {\rm e}^{i\theta(y)}\sigma^{\underline{2}}\bar\zeta(y).
\label{killing_spinor}
\end{eqnarray}
In addition we make the warped metric Ansatz : 
\begin{eqnarray*}
ds^2 = {\rm e}^{2A(y)}\eta_{\mu\nu}dx^\mu dx^\nu + dy^2
\quad
\left(\mu,\nu = 0,1,3\right).
\end{eqnarray*}
The BPS equations can be derived by demanding that the bosonic
configuration should satisfy $\delta_\zeta\psi_m = \delta_\zeta\chi = 0$
for the Killing spinors  $\zeta(y)$ in 
Eq.(\ref{killing_spinor})\cite{Cvetic:1992bf}.
From the condition $\delta_\zeta\psi_{\mu}=0$ 
we obtain the BPS equation for the warp factor:
\begin{eqnarray}
\dot A = - i\kappa^2{\rm e}^{-i\theta}{\rm e}^{\frac{\kappa^2}{2}K}P_{\rm lc},
\label{BPSeq;warp}
\end{eqnarray}
where a dot denotes a derivative with respect to the extra coordinate $y$.
From the condition $\delta_\zeta\psi_2=0$ we obtain the BPS 
equations for the phase $\theta$ and the modulus $|\zeta_\alpha|$ 
of the Killing spinor : 
\begin{eqnarray}
\dot\theta = - \kappa^2 {\rm Im}\left[\sum_i\dot\phi^i\partial_iK\right],
\label{BPSeq;killing_spinor-phase}
\end{eqnarray}
\begin{eqnarray}
|\dot\zeta_\alpha| = \frac{\dot A}{2}|\zeta_\alpha|.
\label{BPSeq;killing_spinor-mod}
\end{eqnarray}
From the remaining condition $\delta_\zeta\chi^i=0$ we obtain the 
BPS equations for the scalar fields:
\begin{eqnarray}
\dot\phi^i = - i {\rm e}^{i\theta} 
{\rm e}^{\frac{\kappa^2}{2}K} K^{ij*}D_{j*}P_{\rm lc}^*.
\label{BPSeq;matter}
\end{eqnarray}
Eqs.(\ref{BPSeq;warp}), (\ref{BPSeq;killing_spinor-phase}), 
(\ref{BPSeq;killing_spinor-mod}) and 
(\ref{BPSeq;matter}) are the full set of our BPS equations. 

Notice that we can recover results of the global
SUSY \cite{SingleWall} if we take the gravitational
coupling $\kappa$ to zero and identify the superpotential $P_{\rm lc}$ in
SUGRA with the superpotential $P_{\rm gl}$ in SUSY.

\subsection{Deformation of Superpotential
%\label{sec:level2}Second-level heading: Formatting
}
Recently, we found the exact BPS solution in ${\cal N}=1$ SUGRA
sine-Gordon model \cite{Eto:2002ns} by allowing a modification of the
superpotential. The gravitational deformation for the superpotential is
originally introduced in order to keep the periodicity of the model with
the aid of the K\"ahler transformation. In this paper we extend the 
gravitational modification of the superpotential 
to other models in order to obtain exact BPS solutions 
in SUGRA theories even in models without particular symmetry, 
such as the periodicity in sine-Gordon model. 

We first note that SUSY theories are always invariant under 
the following K\"ahler transformations 
\begin{eqnarray}
\left\{
\begin{array}{l}
K(\phi,\phi^*) \rightarrow 
K(\phi,\phi^*) + F(\phi) + F^*(\phi^*),\\
P_{\rm gl}(\phi) \rightarrow P_{\rm gl}(\phi),
\label{Kahler_global}
\end{array}
\right.
\end{eqnarray}
where $F(\phi)$ is a holomorphic function of $\phi^i$. 
Moreover, we also note that no physical difference arises 
if we add a constant $a$ in the superpotential 
$P_{\rm gl}$ in global SUSY theories 
\begin{eqnarray}
\tilde P_{\rm gl} \equiv P_{\rm gl}+a. 
\label{eq:constant-in-superpot}
\end{eqnarray}

On the other hand,  the K\"ahler transformations in 
SUGRA theories 
should accompany the transformations of superpotential 
\begin{eqnarray}
\left\{
\begin{array}{l}
K(\phi,\phi^*) \rightarrow 
K(\phi,\phi^*) + F(\phi) + F^*(\phi^*),\\
P_{\rm lc}(\phi) \rightarrow 
{\rm e}^{-\kappa^2F(\phi)}P_{\rm lc}(\phi), 
\label{Kahler_local}
\end{array}
\right.
\end{eqnarray}
and the Weyl transformations of fermions \cite{Wess:cp} : 
\begin{eqnarray*}
\chi^i \rightarrow 
{\rm e}^{\frac{i\kappa^2}{2}{\rm Im}[F(\phi)]}\chi^i,
\quad
\psi_m \rightarrow 
{\rm e}^{-\frac{i\kappa^2}{2}{\rm Im}[F(\phi)]}\psi_m.
\end{eqnarray*}
We observe that an additive constant $a$ in 
Eq.(\ref{eq:constant-in-superpot}) 
does have a physical implication when coupled to 
gravity in SUGRA, in contrast to the global SUSY theories. 

Since the superpotential of the global SUSY theories 
does not transform under the K\"ahler transformations, 
we need to make a 
gravitational deformation of the superpotential if we wish 
to make the superpotential to be invariant under the K\"ahler 
transformations in SUGRA when it is embedded into SUGRA 
theories. 
Then 
the embedded theories are assured 
to have a smooth limit of vanishing gravitational coupling, 
and their vacua as well as solutions are likely 
to be preserved. 

Now let us define a new holomorphic function $\tilde K$ 
from K\"ahler potential by 
replacing $\phi^*$ with $\phi$ in $K$ : 
\begin{eqnarray*}
\tilde K(\phi) \equiv K(\phi,\phi^*\rightarrow\phi).
\end{eqnarray*}
This new function transforms as $\tilde K \rightarrow \tilde K + 2 F$ 
under the K\"ahler transformation (\ref{Kahler_local}) of SUGRA. 
Then, we will choose the SUGRA superpotential $P_{\rm lc}$ from the global
SUSY superpotential with a gravitational deformation as 
\begin{eqnarray}
P_{\rm lc}(\phi) 
= {\rm e}^{-\frac{\kappa^2}{2}\tilde K(\phi)}
\left(P_{\rm gl}(\phi)+a\right)
={\rm e}^{-\frac{\kappa^2}{2}\tilde K(\phi)}
\tilde P_{\rm gl}(\phi),
\label{DSP}
\end{eqnarray}
where we take into account of the possibility of adding 
a constant $a$ in $P_{\rm gl}$ which is physically 
meaningful only when coupled to SUGRA. 
As for the  K\"ahler potential $K$ we choose 
the same K\"ahler 
potential as the global SUSY. 
With this gravitational deformation, the corrected 
superpotential 
$P_{\rm lc}$ automatically obeys the transformation 
law (\ref{Kahler_local}) of 
the SUGRA theory, as a 
consequence of the globally SUSY K\"ahler transformation 
(\ref{Kahler_global}). 
%\footnote{
We shall show in Appendix 
that the above gravitational deformation 
is the unique possibility if we require that the BPS 
equations for matter scalars in the SUGRA theories 
should be identical to those in the global SUSY theories 
assuming three conditions : 
minimal kinetic term (or nonlinear sigma model 
that can be transformed to minimal kinetic term), 
only single scalar field has nontrivial configuration in BPS 
solutions, and is real. 

One can shift the gravitational 
deformation of the superpotential to that of 
the K\"ahler potential 
by making another K\"ahler transformation 
(\ref{Kahler_local}) with $F = - \frac{1}{2}\tilde K$. 
It is interesting to recall that the SUGRA theories in five dimensions 
requires a gravitational deformation of target space manifolds 
of hypermultiplets from hyper-K\"ahler to quaternionic 
K\"ahler manifolds \cite{BaggerWitten}. 
To find out the necessary gravitational deformations 
with a smooth limit of the vanishing gravitational coupling 
has been a challenge for some time \cite{Arai:2002ph,  Eto:2003ut}, 
\cite{SUGRAwall, Beh-Dall}, \cite{Lazaroiu}, 
\cite{Galicki, IvanovValent}. 
In this respect, it is quite natural that a coupling to SUGRA 
in four dimensions also accompanies gravitational deformations 
of the K\"ahelr potential 
or/and the superpotential, since only a combination of K\"ahler potential 
and the superpotential has an invariant meaning in ${\cal N}=1$ 
SUGRA in four-dimensions \cite{Wess:cp}. 
Although scalar fields in SUGRA are generically 
a nonlinear sigma model in this sense, 
we choose here to give gravitational deformations 
to the superpotential 
rather than to the K\"ahler potential. 
%the superpotential $P_{\rm lc}$ with the special gravitational deformation 
%in Eq.(\ref{DSP}) reduces to $P_{\rm gl}$. Instead, the K\"ahler potential 
%take over the gravitational deformation.}.

Notice that this deformed superpotential $P_{\rm lc}$ 
reduces to the global SUSY superpotential $P_{\rm gl}$ 
when we turn off
the gravitational coupling $\kappa$.
As we will see shortly, this special choice of the 
superpotential in 
Eq.(\ref{DSP}) allows us to obtain 
%a useful prescription for 
exact BPS solutions in ${\cal N}=1$ SUGRA. 

The covariant derivative (\ref{CD_SUGRA}) in target 
space acting on 
the gravitationally deformed 
superpotential (\ref{DSP}) becomes  
\begin{eqnarray}
D_iP_{\rm lc} 
= {\rm e}^{-\frac{\kappa^2}{2}\tilde K}
  \left[
   \partial_iP_{\rm gl} 
+ \kappa^2\tilde P_{\rm gl}\partial_i
\left(K - \frac{1}{2}\tilde K\right)
  \right]. 
\label{MCD_SUGRA}
\end{eqnarray}
and the scalar potential (\ref{scalar_pot}) is of the form:
%\begin{widetext}
\begin{eqnarray}
V_{\rm lc} 
\!\!&=&\!\! {\rm e}^{\kappa^2(K-{\rm Re}[\tilde K])} 
  \Biggl[
   K^{ij*}
  \left\{
   \partial_iP_{\rm gl} + \kappa^2\tilde P_{\rm gl}\partial_i
\left(K - \frac{1}{2}\tilde K\right)
  \right\}
\nonumber \\
\!\!&\times&\!\!\!
  \left\{
   \partial_{j*}P_{\rm gl}^* 
   + \kappa^2\tilde P_{\rm gl}^*\partial_{j*}\!\!\!
\left(K - \frac{1}{2}\tilde K^*\right)
  \right\}
   -3\kappa^2|\tilde P_{\rm gl}|^2
  \Biggr].\label{M_scalar_pot}
\end{eqnarray}
%\end{widetext}
Notice that the covariant derivative (\ref{MCD_SUGRA}) 
is covariant
and the scalar potential (\ref{M_scalar_pot}) is 
invariant under 
the SUGRA K\"ahler transformation (\ref{Kahler_local}), 
as a 
consequence of covariance (invariance) under 
the globally SUSY K\"ahler transformation 
(\ref{Kahler_global}).

 For simplicity, we concentrate on the model with the 
 minimal K\"ahler
potential $\displaystyle K= \sum_i|\phi^i|^2$ 
in what follows.
Then, the covariant derivative (\ref{MCD_SUGRA}) 
and the scalar
potential (\ref{M_scalar_pot}) become : 
%\begin{widetext}
\begin{eqnarray}
D_iP_{\rm lc} 
%&
=
%& 
{\rm e}^{-\frac{\kappa^2}{2}\sum_i\phi^{i2}}
  \left[
 \partial_iP_{\rm gl} - 2i\kappa^2 {\rm Im}[\phi^i]
\tilde P_{\rm gl}
  \right],\label{DP:minimal}
%\\
\end{eqnarray}
\begin{eqnarray}
\!\!\!V_{\rm lc}
\!&=&\!
 {\rm e}^{-2\kappa^2\sum_i{\rm Im}[\phi^i]^2} 
\nonumber \\
&\times&\!\!\!  \left[
  \sum_i
     \left|
   \partial_iP_{\rm gl} - 2i\kappa^2 {\rm Im}[\phi^i] 
   \tilde P_{\rm gl}
  \right|^2
     -3\kappa^2|\tilde P_{\rm gl}|^2
  \right].
\end{eqnarray}
%\end{widetext}
The gravitationally deformed superpotential and scalar potential 
takes a simpler form if we restrict ourselves to a section of the
bosonic fields space where all scalar fields $\phi^i$ take only 
real values (real section). 
This is often the case when we consider 
models whose vacua are distributed on the real axes in the complex 
fields space, and BPS domain walls which interpolate these vacua. 
In fact, we will deal with such models in the following section.
For that real section of field space, the scalar potential reduces to 
\begin{eqnarray}
V_{\rm lc} = \sum_i\left(\partial_iP_{\rm gl}\right)^2 
- 3\kappa^2(P_{\rm gl}+a)^2.
\label{eq:scalar-pot-real-sec}
\end{eqnarray}
Here we assume that couplings in the superpotential are real parameters 
and $\phi^i=\phi^{i*}$. Then, the superpotential is essentially a real 
function of the real field $\phi^i$. 
Notice that this type of the 
scalar potential often arises in a restricted section of 
field space of the $D$ dimensional 
SUGRA \cite{Skenderis:1999mm, Cvetic:2002su} 
and ensures stable AdS vacua, 
if the superpotential has at 
least a stationary point \cite{Boucher}, \cite{Townsend:iu}. 
The first term in Eq.(\ref{eq:scalar-pot-real-sec}) 
%completely 
corresponds to the term which comes from the 
$F$-term in global SUSY, whereas the second term expresses a 
gravitational correction. 
The first term vanishes at SUSY 
vacua and generally the second term is non-vanishing. 
Therefore, the vacuum 
becomes AdS (or flat if $a$ happens to cancel with the 
$-P_{\rm gl}(\langle\phi\rangle)$ spacetime, 
since the second term gives a 
negative cosmological constant. 

One of the most important points is that the locations of the
SUSY vacua for the superpotential $P_{\rm lc}$ given in Eq.(\ref{DSP}) are
not changed from the global SUSY model with 
superpotential $P_{\rm gl}$. The SUSY vacua are determined from the
``$F$-term'' condition in SUGRA \cite{Wess:cp} :
\begin{eqnarray*}
%{\rm e}^{\frac{\kappa^2}{2}K}K^{ij*}
D_{i}P_{\rm lc} = 0.
\end{eqnarray*}
Because of Eq.(\ref{DP:minimal}) and our choice of 
the superpotential $P_{\rm lc}$ in Eq.(\ref{DSP}) with the minimal
K\"ahler potential, 
this condition agrees exactly with that of global SUSY 
for the real section of the field space. 
Hence, the SUGRA theory with our gravitationally deformed 
superpotential have the SUSY vacua that are precisely identical to the 
SUSY vacua of the global SUSY theory at the 
stationary points of %the global SUSY superpotential 
$P_{\rm gl}$.

The BPS equation (\ref{BPSeq;killing_spinor-phase}) 
for the phase $\theta(y)$ of Killing spinor defined 
in Eq.(\ref{killing_spinor}) implies that the phase 
should be independent of $y$ for the BPS solution with real 
scalar field configurations. 
The reality of scalar fields is consistent with the 
BPS equation for matter fields (\ref{BPSeq;matter}) 
only if $\theta=\pm \pi/2+2n\pi, \; n\in {\bf Z}$ : 
\begin{eqnarray}
\dot\phi = {\rm e}^{i\left(\theta - \frac{\pi}{2}\right)}
\partial_iP_{\rm gl}(\phi), \label{BPSeq;matter_global}
\end{eqnarray}
which is exactly identical to the BPS equation in global SUSY 
theories with the globally SUSY superpotential $P_{\rm gl}$.  
We will refer the case of $\theta = {\pi}/{2}+2n\pi$ 
($\theta = -{\pi}/{2}+2n\pi$) as the 
BPS (anti-BPS) solutions. 
Therefore, we can automatically obtain exact BPS solutions in SUGRA, if
we choose the superpotential according to Eq.(\ref{DSP}).
The warp factor and the Killing spinor are also obtained from the
other BPS equations (\ref{BPSeq;warp}) and (\ref{BPSeq;killing_spinor-mod}) 
\begin{eqnarray}
\dot A &=& \kappa^2 {\rm e}^{-i\left(\theta + \frac{\pi}{2}\right)} P_{\rm gl},
\label{BPSeq;warp_global}\\
\zeta_\alpha &=& {\rm e}^{\frac{i}{2}\left(\theta + \frac{\pi}{2}\right)}
{\rm e}^{\frac{A}{2}}\times
\left(
\begin{array}{c}
\epsilon_1\\
\epsilon_2
\end{array}
\right),\label{BPSeq;killing_spinor_global}
\end{eqnarray}
where $\epsilon_{1(2)}$ represents a constant Grassmann parameter 
corresponding to the two conserved SUSY directions. 

Notice that the energy density of the BPS domain wall obtained here 
is precisely identical to the one in the global SUSY model 
which is given by the
topological charge \cite{SingleWall} 
\begin{equation}
Z = 2|\Delta P_{\rm gl}|. 
\label{eq:topological-charge}
\end{equation}
We shall illustrate this point 
for concrete examples in the following sections. 

\section{Exactly solvable Examples}
\label{sc:solvable-examples}
\subsection{Double Well Model}

\begin{figure*}[t]
\includegraphics[width=15cm]{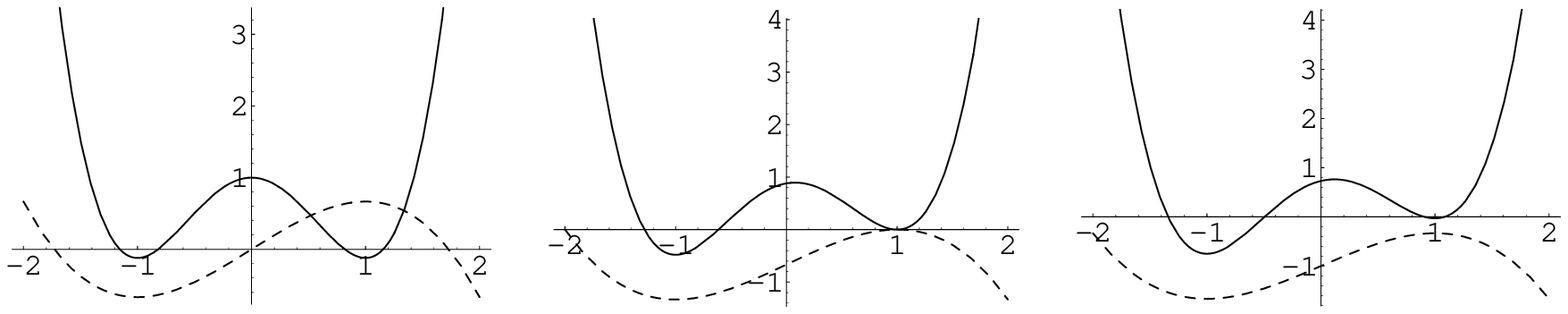}
\caption{The scalar potential $V_{\rm lc}$ (solid line) and the global
 SUSY superpotential $P_{\rm gl}$
 (broken line) as a function of ${\rm Re}[\phi]$. 
Parameters are taken to be 
 $\kappa=0.3,g=1$ and $\Lambda=1$. 
From left to right, $a$ is $(0,-2/3,-1)$, corresponding to 
IR-IR, IR-flat, IR-UV behaviors.}
\label{pot:AdS/CFT}\ \\
\includegraphics[width=15cm]{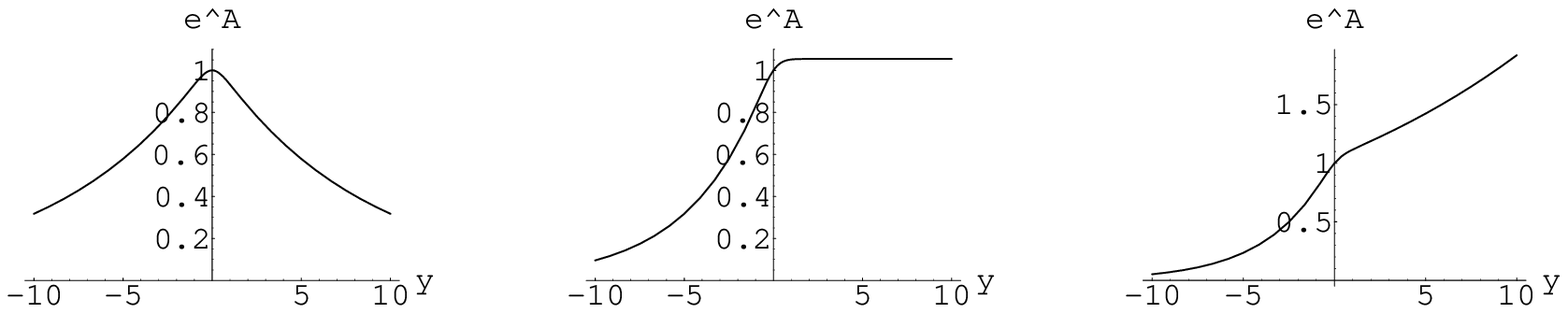}
\caption{The warp factor ${\rm e}^{A}$ as a function of $y$. 
Parameters are taken to be 
 $\kappa=0.3,g=1$ and $\Lambda=1$. 
From left to right, $a$ is $(0,-2/3,-1)$, corresponding to 
IR-IR, IR-flat, IR-UV behaviors.}
\label{warp:AdS/CFT}
\end{figure*}

To obtain a stable BPS domain wall in the global SUSY model, 
there must be at least two isolated SUSY vacua. 
Assuming the minimal kinetic term, 
one of the simplest superpotentials which give such vacua in 
global SUSY theories is the $\phi^3$ type : 
\begin{eqnarray}
P_{\rm gl} = \Lambda^2\phi - \frac{g}{3}\phi^3,
\label{SP_sin_gl}
\end{eqnarray}
where $g,\Lambda$ are both real positive coupling constants.
The SUSY vacua are given as stationary points of the 
superpotential :
\begin{eqnarray}
\langle\phi\rangle 
= \left( \frac{\Lambda}{\sqrt{g}}, 
-\frac{\Lambda}{\sqrt{g}}\right),\quad
\langle P_{\rm gl} \rangle 
= \left(\frac{2}{3}\frac{\Lambda^3}{\sqrt{g}},
-\frac{2}{3}\frac{\Lambda^3}{\sqrt{g}}\right),
\label{vacua_phi^3}
\end{eqnarray}
whose vacuum energy density vanishes as a consequence of 
SUSY. 
The BPS domain wall which interpolates these two vacua is the solution of
the BPS equation (\ref{BPSeq;matter_global}):
\begin{eqnarray}
\phi(y) = {\rm e}^{i\left(\theta - \frac{\pi}{2}\right)}\frac{\Lambda}{\sqrt{g}}\tanh\Lambda\sqrt{g}(y-y_0),
\label{BPSsol;phi^3}
\end{eqnarray}
where $\theta=\pm\pi/2$ and 
$y_0$ is a collective coordinate which corresponds to the wall
position.
The energy density of the BPS wall solution is given by the 
topological charge in Eq.(\ref{eq:topological-charge}) 
\begin{eqnarray}
{\cal E} 
 = Z \equiv 
2 \left|\Delta P_{\rm gl}\right| 
 = \frac{8}{3}\frac{\Lambda^3}{\sqrt{g}}.
\label{Enegy_phi^3}
\end{eqnarray}

This global SUSY model has been studied by coupling to SUGRA 
without any gravitational deformations of the superpotential 
\cite{Cvetic:1992bf}. 
It may be instructive to compare two SUGRA theories : 
one with our gravitationally 
deformed superpotential $P_{\rm lc}$ given by the prescription (\ref{DSP}) 
and the other with the superpotential $P_{\rm gl}$ (\ref{SP_sin_gl}) 
without gravitational deformations 
inserted in place of the $P_{\rm lc}$ in the SUGRA Lagrangian 
(\ref{main_lagrangian}) and (\ref{scalar_pot}). 
For the latter choice, the SUSY
vacuum condition $D_\phi P_{\rm lc}=0$ in the SUGRA theory 
gives two isolated SUSY vacua : 
\begin{eqnarray*}
\langle\phi\rangle 
= \left( \sqrt{-\alpha+\sqrt{\alpha^2 + \beta}}, 
-\sqrt{-\alpha+\sqrt{\alpha^2 + \beta}}\right),
\end{eqnarray*}
where $\alpha(g,\Lambda;\kappa) \equiv 
\frac{3(g-\kappa^2\Lambda^2)}{2g\kappa^2}$ and $\beta(g,\Lambda;\kappa)
\equiv \frac{3\Lambda^2}{g\kappa^2}$. 
These reduce to Eq.(\ref{vacua_phi^3}) when we turn off the
gravitational coupling $\kappa$. 
However, it is generally difficult to obtain an 
exact BPS solution, since these vacua have a nontrivial 
dependence on the gravitational coupling $\kappa$. 
The wall configuration has been studied numerically and 
it is reasonably compelling that 
the wall solution should exist \cite{Cvetic:1992bf}, 
although an explicit demonstration of the solution was difficult.

On the contrary, according to our prescription of gravitational 
deformations in Eq.(\ref{DSP}) 
%which leads to an exact solution in SUGRA, 
we should choose the superpotential in SUGRA as 
\begin{eqnarray*}
P_{\rm lc} = {\rm e}^{-\frac{\kappa^2}{2}\phi^2}
\left(\Lambda^2\phi - \frac{g}{3}\phi^3+a\right).
\end{eqnarray*}
The BPS equations (\ref{BPSeq;warp}), 
(\ref{BPSeq;killing_spinor-phase}), 
(\ref{BPSeq;killing_spinor-mod}) and 
(\ref{BPSeq;matter}) in SUGRA 
with this modified superpotential give the same vacua as
Eq.(\ref{vacua_phi^3}) and precisely the same exact 
BPS solution 
 in Eq.(\ref{BPSsol;phi^3}) which interpolates 
these two vacua. 
The vacuum energy densities in these two SUSY vacua no 
longer vanish but are negative 
\begin{eqnarray*}
V_{vac} = - {3\kappa^2}\left(\epsilon(y)
\frac{2}{3} 
\frac{\Lambda^3}{\sqrt{g}} 
+a\right)^2. 
%\label{Vac_Ene_phi^3} 
\end{eqnarray*}
Therefore the BPS domain wall in Eq.(\ref{BPSsol;phi^3}) 
interpolates two AdS vacua with decreasing warp factor 
asymptotically to both infinities, if $|a|<\frac{2}{3} 
\frac{\Lambda^3}{\sqrt{g}}$. 
This is phenomenologically desirable situation corresponding 
to IR fixed points in both infinities with respect to 
AdS/CFT correspondence \cite{AdS/CFT}. 
If $a=-\frac{2}{3} \frac{\Lambda^3}{\sqrt{g}}$ 
 ($+\frac{2}{3}\frac{\Lambda^3}{\sqrt{g}}$), 
positive (negative) asymptotic infinity is flat space, 
whereas the other infinity is AdS space and the 
warp factor is exponentially decreasing. 
If $a<-\frac{2}{3} \frac{\Lambda^3}{\sqrt{g}}$ 
 ($a>\frac{2}{3} \frac{\Lambda^3}{\sqrt{g}}$), 
both asymptotic infinities are AdS spaces, 
but the warp factor is exponentially 
increasing at positive (negative) 
asymptotic infinity 
and is exponentially decreasing at 
negative (positive) asymptotic infinity. 

We can also obtain exact BPS solutions for the warp factor 
from Eq.(\ref{BPSeq;warp_global}) 
\begin{eqnarray*}
A &=& -\kappa^2 \bigg[
a y \\&+& \frac{2\Lambda^2}{3g}\left(
\log\cosh\Lambda\sqrt{g}(y-y_0)
      + 
\frac{\tanh^2\Lambda\sqrt{g}(y-y_0)}{4}       
\right)
\bigg].
\end{eqnarray*}
The scalar potential and the global SUSY superpotential with the parameters 
$a=(0,-2/3,-1){\Lambda^3}/{\sqrt{g}}$ are
shown in FIG.\ref{pot:AdS/CFT}.
In FIG.\ref{warp:AdS/CFT} the profiles of the warp factor are shown.
The Killing spinor $\zeta_\alpha$ is obtained 
by plugging this warp factor into 
Eqs.(\ref{BPSeq;killing_spinor_global}).

Notice that the
energy density of the BPS domain wall is just the 
same as that of the
global SUSY model in Eq.(\ref{Enegy_phi^3}). 
In the case of no gravitational deformations 
for the superpotential, 
it was shown that the energy density of the 
BPS domain wall in SUGRA generally differs 
from that of the global 
SUSY by a factor arising from the K\"ahler potential  
\cite{Cvetic:1992bf} 
\begin{eqnarray*}
{\cal E} = 
2\left|{\rm e}^{\frac{\kappa^2}{2}K}|
\Delta P_{\rm lc}|\right|
\not = Z.
\end{eqnarray*}
In our model, 
this factor from the K\"ahler potential is 
absorbed in the gravitationally deformed superpotential 
in Eq.(\ref{DSP}), so that the 
energy density in SUGRA is precisely the same as that 
in global SUSY.

\subsection{Sine-Gordon Model}

In Ref.\cite{Eto:2002ns} we found the exact BPS 
solution for the modified sine-Gordon model 
with the superpotential given here except the possible 
additive constant $a$. 
Here we also add a possibility of this additive constant 
in superpotential :
\begin{eqnarray*}
P_{\rm lc} = 
{\rm e}^{-\frac{\kappa^2}{2}\phi^2}
\left(
\frac{\Lambda^3}{g^2}\sin\frac{g}{\Lambda}\phi 
+ a \right).
\end{eqnarray*}
The SUSY vacua is periodically distributed on the real axis 
in the complex $\phi$
plane:
\begin{eqnarray*}
\langle \phi \rangle 
= \frac{\Lambda}{g}\left(\frac{\pi}{2} + n\pi\right),\quad
(n\in\mathbb Z).
\end{eqnarray*}
The BPS solution which interpolates any two adjacent vacua is 
of the form:
\begin{eqnarray*}
\phi &=& \frac{\Lambda}{g}
         \left[
  (-1)^n\left\{2\tan^{-1}{\rm e}^{\pm \Lambda(y-y_0)}-\frac{\pi}{2}\right\}
	  + n\pi
	 \right],\\
A &=& -\kappa^2 \left[a y
 + \frac{\Lambda^2}{g^2}\log\cosh\Lambda(y-y_0)\right].
\end{eqnarray*}
More details about this model are given in Ref.\cite{Eto:2002ns}.

\subsection{Two Walls Model}

An interesting global SUSY model in four dimensions 
has been found which allows 
two domain walls as an exact BPS solution 
\cite{Shifman:1997wg, GK}. 
The model consists of two chiral superfields 
$\Phi$ and $X$ whose lowest components are denoted by 
$\phi$ and $\chi$, respectively. 

Let us consider the minimal kinetic term given by 
the minimal K\"ahler potential : $K = |\phi|^2 + |\chi|^2$.
The simplest (trivial) model admitting two walls in global SUSY 
consists of two 
decoupled double-well model:
\begin{eqnarray}
P_{\rm gl} = \Lambda_\phi^2\phi - \frac{g_\phi}{3}\phi^3 
       + \Lambda_\chi^2\chi - \frac{g_\chi}{3}\chi^3,
\label{eq:decoupled-double-well}
\end{eqnarray}
where all the couplings $\Lambda_\phi, \Lambda_\chi, g_\phi, g_\chi$ 
are assumed to be real positive.
This superpotential gives four isolated SUSY vacua:
\begin{eqnarray*}
\left(
\begin{array}{c}
\phi\\
\chi
\end{array}
\right)
= 
\left(
\begin{array}{c}
\pm \Lambda_\phi/\sqrt{g_\phi}\\
\pm \Lambda_\chi/\sqrt{g_\chi}
\end{array}
\right),
\left(
\begin{array}{c}
\pm \Lambda_\phi/\sqrt{g_\phi}\\
\mp \Lambda_\chi/\sqrt{g_\chi}
\end{array}
\right).
\end{eqnarray*}
Since $\phi$ and $\chi$ are decoupled, the exact 
BPS solution is
a superposition of that of each double-well model:
\begin{eqnarray}
\phi &=& \epsilon_\phi\frac{\Lambda_\phi}{\sqrt{g_{\phi}}}
\tanh\Lambda_\phi(y-y_\phi),
\label{two_wall_simple:phi}\\
\chi &=& \epsilon_\chi\frac{\Lambda_\chi}{\sqrt{g_{\chi}}}
\tanh\Lambda_\chi(y-y_\chi),
\label{two_wall_simple:chi}
\end{eqnarray}
where $\epsilon_{\phi(\chi)}$ is $\pm 1$.
This solution has two collective coordinates: 
the center of the mass
$y_{cm} = \frac{y_\phi + y_\chi}{2}$ and the relative 
distance between
the two walls $R = |y_\phi - y_\chi|$.

In general, this superposition principle does not hold 
in gravity theory. 
Namely, the superposition of the individual solutions 
is not a solution. 
That is because two scalar fields $\phi,\chi$ are coupled 
via gravity even 
if these are decoupled in the superpotential. 
So the superposition of the 
solutions of the individual models does not satisfy 
the equations
of motion when they are coupled to gravity theory. 
We can see this gravitational interaction 
in the BPS equation 
(\ref{BPSeq;matter}) if we use the superpotential 
(\ref{eq:decoupled-double-well}) 
of the global SUSY model 
(without the gravitational deformations) 
inserted into the SUGRA superpotential $P_{\rm lc}$. 
Two fields are coupled through the K\"ahler
potential. 
It is then difficult to obtain the exact BPS solutions 
for two walls in SUGRA. 

If we choose the superpotential according to our prescription 
(\ref{DSP}), 
\begin{eqnarray*}
P_{\rm lc} = {\rm e}^{-\frac{\kappa^2}{2}(\phi^2+\chi^2)}
         \left[
	  \Lambda_\phi^2\phi - \frac{g_\phi}{3}\phi^3 
	  + \Lambda_\chi^2\chi - \frac{g_\chi}{3}\chi^3 + a
	 \right],
\end{eqnarray*}
two fields behave as if they are effectively decoupled 
even in the presence of gravity. 
Then we obtain the exact solution for scalar fields 
identical to 
Eqs.(\ref{two_wall_simple:phi}) and
(\ref{two_wall_simple:chi}). 
The BPS solution of the warp factor is of the form:
\begin{eqnarray*}
 A \!&=&\! 
     -\kappa^2 a y\\
     &-&\!\! \frac{\kappa^2\Lambda_\phi^2}{\sqrt{g_\phi}}
     \left[
  \frac{2}{3}\log\cosh\Lambda_\phi(y-y_\phi)
 + \frac{\tanh^2\Lambda_\phi(y-y_\phi)}{6} 
     \right]\\
     &-&\!\! \frac{\kappa^2\Lambda_\chi^2}{\sqrt{g_\chi}}
     \left[
  \frac{2}{3}\log\cosh\Lambda_\chi(y-y_\chi)
 + \frac{\tanh^2\Lambda_\chi(y-y_\chi)}{6} 
     \right].
\end{eqnarray*}

Next we turn our attention to the more interesting model. 
In Ref.\cite{Shifman:1997wg, GK}, an integral of motion was 
constructed for a global SUSY model with two chiral scalar fields 
 with the superpotential 
\begin{eqnarray*}
P_{\rm gl} = \frac{m^2}{\lambda}\phi 
- \frac{\lambda}{3}\phi^3 - \alpha\phi\chi^2,
\end{eqnarray*}
where all the coupling constants are real and positive. 
This model has the following four 
isolated SUSY vacua : 
\begin{eqnarray*}
\left(
\begin{array}{c}
\phi\\
\chi
\end{array}
\right)
= 
\left(
\begin{array}{c}
\pm m/\lambda\\
0
\end{array}
\right),
\left(
\begin{array}{c}
0\\
\pm m/\sqrt{\alpha\lambda}
\end{array}
\right).
\end{eqnarray*}
The BPS equations for matter fields can be cast into 
dimensionless forms :
\begin{eqnarray}
{d f \over d(my)} = 1 - f^2 - h^2,\quad
{d h \over d(my)} = - \frac{2}{\rho} fh,
\label{eq:two-field-BPS}
\end{eqnarray}
where $f,h$ are dimensionless fields defined as 
$f \equiv (\lambda/m) \phi$
and $h \equiv (\sqrt{\alpha\lambda}/m) \chi$, $\rho$ 
is defined as 
$\rho \equiv \lambda/\alpha$. 
By taking the ratio of Eqs.(\ref{eq:two-field-BPS}) to 
eliminate the  $u=my$ dependence, 
we can integrate once 
to obtain an implicit solution giving a relation between $f$ and $h$ 
\begin{eqnarray*}
f^2 = 1 - \frac{\rho}{\rho-2}h^2 - Ch^\rho,
\end{eqnarray*}
where $C$ is an integration constant which corresponds 
to a collective
coordinate related with the relative distance 
between two walls. 
Supplemented by one of the above Eqs.(\ref{eq:two-field-BPS}), 
one can explicitly obtain the complete solution. 

 For the special case where $\rho = 4$ these BPS equations 
(\ref{eq:two-field-BPS}) are explicitly solvable. 
To clarify the physical meaning of 
the relative distance between walls, 
it is more convenient to 
convert the integration constant $C$ into the following 
parameter \cite{GK}
$
%t
%\begin{eqnarray}
t\equiv 1/\sqrt{C+1}
$. %\end{eqnarray}
The exact BPS two wall solution is given with a center of 
position $y_0$ and another moduli parameter $t$ 
by 
\begin{eqnarray}
\phi &=& \frac{m}{\lambda}
\frac{\sinh m(y-y_0)}{\cosh m(y-y_0)+t},
\label{2wall_phi}\\
\chi &=& \pm \frac{m}{2\lambda}
\sqrt{\frac{t}{\cosh m(y-y_0)+t}},
\label{2wall_chi}
\end{eqnarray}
where the moduli parameter $t$ can take values $0<t<\infty$ 
corresponding to $\infty > C > -1$. 
For $1<t$, we can convert it to $t=\cosh s$ which can be 
interpreted as the $m$ times distance between walls. 
We obtain 
\begin{eqnarray*}
\phi &=& \frac{m}{2\lambda}
\left(\tanh {u-s \over 2}
+\tanh {u+s \over 2}\right),
\label{2wall_phi2}\\
\chi &=& \pm \frac{m}{2\lambda}
\sqrt{\frac{1}{2}\left(1-
\tanh {u-s \over 2}
\tanh {u+s \over 2}\right)},
\label{2wall_chi2}
\end{eqnarray*}
\begin{eqnarray*}
u\equiv m(y-y_0), 
\end{eqnarray*}
and interpret the moduli parameter to be 
the distance between two walls. 

As we have seen, these solutions remain to be 
BPS solutions for the SUGRA theories, if we choose the 
superpotential : 
\begin{eqnarray*}
P_{\rm lc} = {\rm e}^{-\frac{\kappa^2}{2}(\phi^2+\chi^2)}
       \left(
\frac{m^2}{\lambda}\phi - \frac{\lambda}{3}\phi^3 
- \alpha\phi\chi^2
+a\right).
\end{eqnarray*}
\begin{figure}[htb]
%\vspace{-1cm}
\includegraphics[width=8cm]{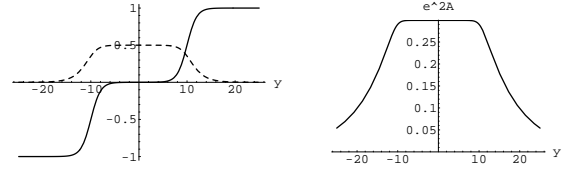}
\caption{Two walls solution for the scalar fileds and the warp factor is shown. 
The parameters are taken to be $\kappa=0.3,m=1,\lambda=1$ 
and $a=0$. 
We take $s=10$.}
\label{warp_2wall}
\end{figure}
Plugging Eqs.(\ref{2wall_phi}) and 
(\ref{2wall_chi}) into Eq.(\ref{BPSeq;warp_global}),
we obtain the warp factor explicitly : 
%\begin{widetext}
\begin{eqnarray*}
\!\! A \!\!\!&=&\!\!\! 
- \kappa^2\Biggr[
{a \over m} u \\
&\!\!\!+&\!\!\!\! \frac{m^2}{3\lambda^2}\bigg(\!\!
2\log\left(\cosh u + t\right)
\!+\! \frac{t}{\cosh u  + t}
\!+\! \frac{t^2-1}{2(\cosh u  + t)^2}\!\!
\bigg)\!\Biggr]\!,
\end{eqnarray*}
%\end{widetext}
where an integration constant is suppressed. 
The scalar fields profile and the warp factor are shown in
FIG.\ref{warp_2wall} for $a=0$. 
The case of $a=0$ gives an interesting example for a flat 
bulk between two walls where graviton is confined 
resulting in a flat wave function. 

In the multi-wall solutions with moduli, it is interesting 
to explore dynamics of walls. 
It have been revealed that there is a bound state of walls if one 
introduces winding number into the model considered here 
\cite{SakaiSugisaka}, 
and nonlinear sigma model of moduli fields have also been worked out 
\cite{AGMT}. 
It is an interesting future problem to examine such dynamical 
issues of multi-walls in SUGRA models. 

\subsection{Runaway Vacuum Model}

We have previously studied ${\cal N}=1$ SUSY nonlinear 
sigma models to obtain BPS walls and junction solutions 
\cite{NNS2}. 
Let us consider simplest of these models 
%to obtain wall solutions in nonlinear sigma models 
%in SUGRA theories : 
\begin{eqnarray}
K = \frac{\Lambda^2}{g^2}
\left|\tanh^{-1}\frac{g}{\Lambda}\phi\right|^2, 
\quad
P_{\rm gl} = \frac{\Lambda^2}{g}\phi,
\label{eq:NLSM}
\end{eqnarray}
where $\Lambda,g$ are real positive parameters. 
The K\"ahler metric of this model 
is a product of holomorphic 
and anti-holomorphic part, which is called 
holomorphically factorizable and 
can be transformed into a linear sigma model 
by a holomorphic reparametrization of field \cite{NNS2} 
\begin{eqnarray*}
\phi = \frac{\Lambda}{g}\tanh\frac{g}{\Lambda}\varphi.
\end{eqnarray*}
In terms of $\varphi$, 
the model reduces to a linear sigma model $K=|\varphi|^2$ 
with the superpotential 
$P_{\rm gl}=
\frac{\Lambda^3}{g^2}\tanh\frac{g}{\Lambda}\varphi$. 
In the linear sigma model, the SUSY vacua occur as 
stationary points of the superpotential which is given 
at $\varphi=\pm \infty$ in this particular model. 
These vacua at infinity of field space are often 
called ``runaway vacua'' and are sometimes discarded. 
However, these vacua should be taken into account in order 
to preserve the number of possible SUSY vacua under the 
holomorphic change of field variables \cite{NNS2}. 
These runaway vacua are brought to a finite 
point in field space of $\phi$ in the nonlinear 
sigma model (\ref{eq:NLSM}), 
which realizes these SUSY vacua as 
singular points of the K\"ahler metric : 
\begin{eqnarray*}
K_{\phi\phi^*} = 
\left|\frac{\Lambda^2}{\Lambda^2-g^2\phi^2}\right|^2.
\end{eqnarray*}
Then, the SUSY vacua are 
\begin{eqnarray}
\langle \phi \rangle = 
\left(\frac{\Lambda}{g},-\frac{\Lambda}{g}\right).
\label{SUSY_vac_ranaway}
\end{eqnarray}
Assuming $\phi=\phi^*$, the BPS equation in this global SUSY 
model reduces to %(choosing $\theta=\pi/2$) 
\begin{eqnarray}
\dot\phi 
 =  K^{\phi\phi^*}\partial_{\phi^*}P_{\rm gl}^*
 =  \frac{\Lambda^2}{g}
 \left(1 - \frac{g^2}{\Lambda^2}\phi^2\right)^2,
\label{BPSeq;runaway}
\end{eqnarray}
and its solution is exactly solvable:
\begin{eqnarray}
y - y_0 = \frac{1}{4\Lambda}
          \left(
\frac{\frac{2g}{\Lambda}\phi}{1-\frac{g^2}{\Lambda^2}\phi^2}
+ \log \frac{1 + \frac{g}{\Lambda}\phi}
{1 - \frac{g}{\Lambda}\phi}
	  \right).\label{BPSwall_runaway}
\end{eqnarray}

Let us couple this model with SUGRA, 
according to the prescription
(\ref{DSP}):
\begin{eqnarray*}
P_{\rm lc} = 
{\rm e}^{-\frac{\kappa^2}{2}\tilde{K}} 
\left( 
\frac{\Lambda^2}{g}\phi+a\right),
\quad
%\end{eqnarray*}
%where the holomorphic function $\tilde K$ is defined by
%\begin{eqnarray*}
\tilde{K} = \frac{\Lambda^2}{g^2}
\left(\tanh^{-1}\frac{g}{\Lambda}\phi\right)^2.
\end{eqnarray*}
The SUSY vacua can be obtained by demanding the SUSY 
invariance in Eq.(\ref{SUSY_transf}). 
More explicitly, the SUSY vacua are given by 
Eqs.(\ref{BPSeq;matter}) 
and (\ref{MCD_SUGRA}) as 
\begin{eqnarray}
0 = {\rm e}^{\frac{\kappa^2}{2}K}
K^{\phi\phi^*}D_{\phi^*}P_{\rm lc}^*, 
\label{eq:vacua-runaway}
\end{eqnarray}
\begin{eqnarray*}
D_{\phi^*}P_{\rm lc}^*  
= {\rm e}^{-\frac{\kappa^2}{2}\tilde K^*}
       \!\!\left[
       \frac{\Lambda^2}{g} 
       \!+\! \frac{2i\kappa^2\!\Lambda\!
       \left(\frac{\Lambda^2}{g}\phi^* \!\!+\! a\right)
       {\rm Im}\!\! \left[\!\tanh^{\!-\!1}
       \!\!\frac{g\phi}{\Lambda}\right]}
       {g^2\left(1 - \frac{g^2\phi^{*2}}{\Lambda^2}\right)}
      \right]\!.
\end{eqnarray*}
Eq.(\ref{eq:vacua-runaway}) reduces to 
$0= K^{\phi\phi^*}\Lambda^2/g$ 
for real field configurations $\phi=\phi^*$. 
Therefore the SUSY vacua is unchanged 
from Eq.(\ref{SUSY_vac_ranaway}) given as the zero 
of the inverse K\"ahler metric $K^{\phi\phi^*}$. 
Moreover, the BPS equation (\ref{BPSeq;runaway}) is 
also unchanged, so
the BPS solution is given in Eq.(\ref{BPSwall_runaway}).
The BPS equation for the warp factor 
(\ref{BPSeq;warp_global}) 
can also be integrated to give an expression in terms of 
the scalar field $\phi$ 
\begin{eqnarray*}
A = -\kappa^2 \left[ay
+\frac{\Lambda^2}
{2g^2\left(1- \frac{g^2}{\Lambda^2}\phi^2\right)}\right].
\end{eqnarray*}
Together with Eq.(\ref{BPSwall_runaway}), it 
implicitly gives $A$ 
as a function of $y$.

\section{Zero Mode Wave Function}
\label{sc:zero-mode}
In this section we study the behavior of the warp 
factor ${\rm e}^{2A(y)}$. 
It is well known that the warp factor is closely related to
the zero mode wave function of the graviton in the 
Kaluza-Klein modes expansion. 
The mode equation for the graviton can be written 
as a Schr\"odinger 
type equation \cite{EMS}--\cite{Bazeia:2003cv}: 
\begin{eqnarray*}
\square_{D-1}h^{\rm TT}_{\mu\nu} 
= [-\partial_z^2 + V(z)]h^{\rm TT}_{\mu\nu},\qquad\\
V(z) = \left(\frac{D-2}{2}\frac{dA}{dz}\right)^2 +
 \frac{D-2}{2}\frac{d^2A}{dz^2},
\end{eqnarray*}
where $D$ denotes the spacetime dimension, 
$z$ is the conformal flat coordinate defined by 
$dy = {\rm e}^{A}dz$ and $h^{\rm TT}_{\mu\nu}$ 
is the transverse traceless part of
the fluctuation defined by 
$\delta g_{\mu\nu} = 
{\rm e}^{-\frac{D-6}{2}A}h^{\rm TT}_{\mu\nu}$.
The zero mode wave function of the above 
Schr\"odinger equation is given by 
$h^{\rm TT}_{\mu\nu} = {\rm e}^{\frac{D-2}{2}A(z)}$, 
which implies that the graviton zero mode wave function 
is identical to the warp factor : 
$\delta g_{\mu\nu} = {\rm e}^{2A(y)}$. 
Hence, the behavior of the warp factor is directly 
related to the normalizability of the graviton zero mode 
which is very important for phenomenology.

Next we derive a general property of the warp factor 
for our model. 
Using Eqs.(\ref{BPSeq;matter_global}) and 
(\ref{BPSeq;warp_global}), we find 
\begin{eqnarray*}
\ddot A = - \kappa^2 \sum_i(\dot\phi^i)^2.
\end{eqnarray*}
This implies that the warp factor has at most one 
stationary (maximum) point.  
The physical reason behind this fact should be that 
 matter scalar fields produce only positive 
energy density. 

We also find from Eq.(\ref{BPSeq;warp_global}) 
that the warp factor $A$ has a maximum 
only at the point where the superpotential $P_{\rm gl}$ 
vanishes along the wall trajectory. 
Only such a wall can have the localized graviton 
zero mode around the wall.

If the additive constant 
in the global SUSY superpotential $P_{\rm gl}$ 
(\ref{DSP}) vanishes $a=0$,  
the warp factor $A(y)$ in all the above examples 
are $Z_2$ symmetric under the reflection 
$y-y_0\rightarrow -(y-y_0)$ 
around the center of the wall. 
Therefore the graviton zero 
mode wave function is normalizable 
and we obtain a localized graviton zero mode 
around the wall. 

On the contrary, we have $Z_2$ asymmetric warp factors 
around the wall, if we have non-vanishing additive constant 
$a$. 
It is interesting to remember that 
the constant term in the superpotential has no 
physical effects in global SUSY models. 
In SUGRA theories, however, the constant term $a$ 
produces a dramatic change. 
The BPS equation for the matter in 
 Eq.(\ref{BPSeq;matter_global}) is identical to 
that of the global SUSY, even if we add 
a constant term with the superpotential $P_{\rm gl}$. 
On the other hand the BPS equation for
the warp factor (\ref{BPSeq;warp_global}) is affected 
by the additive constant $a$ 
by controlling the SUSY 
vacuum energy density. 
There is a critical value of the constant $a$ beyond which 
the graviton zero mode ceases to be normalizable : 
$a = - \langle P_{\rm gl} \rangle$. 
At this critical value of the constant $a$, 
the energy density of one side of asymptotic region 
(vacuum) vanishes and the metric in this 
asymptotic infinity reduces to the flat metric. 
Thus we find three different asymptotic behaviors 
of warp factors \cite{Arai:2002ph} 
in the context of AdS/CFT correspondence \cite{AdS/CFT} : 
IR-IR (exponentially decreasing in both infinities), 
IR-flat (exponentially decreasing in one side and 
flat in the other), and IR-UV 
(exponentially decreasing in one side and increasing 
in the other).

\section{Conclusion}
\label{sc:conclusion}
We proposed a prescription for a gravitational deformation 
of the superpotential (or K\"ahler potential) in embedding 
global SUSY models into SUGRA theories. 
This is natural from the viewpoint of the K\"ahler
transformation and gives us exact BPS solutions in SUGRA 
theories from exact solutions in global SUSY models, 
provided scalar field configurations are real. 
Thanks to the gravitational deformations of the 
superpotential, the SUSY vacua in SUGRA theories 
are identical to those in the global SUSY models. 
The domain wall solutions which interpolate 
these vacua are also identical to those of the SUSY models. 

The spacetime distortion by the domain wall 
represented by the warp factor 
can be also obtained by solving the BPS 
equation for the warp factor, which is independent 
of the BPS equation for the matter scalar 
fields. 
The Killing spinor is also obtained in terms of 
the warp factor.

A constant term in the superpotential in global SUSY 
theories has no physical effects. 
However, this constant term gives physical effects 
when coupled to gravity. 
Naturally our gravitationally deformed superpotential 
in SUGRA theories has this freedom of choosing an 
additive constant. 
This constant term has no effect on the scalar field 
configurations, but has significant effects on 
the warp factor. 
Namely, the asymptotic behavior of the warp factor 
is different as the constant term crosses a critical value. 
At the same time, 
the graviton 
zero mode on the domain wall is localized 
for smaller values of the constant, 
whereas it becomes non normalizable 
if the constant is outside of the critical value. 
%When two or more walls exist, the constant superpotential 
%controls not only the localization of the zero mode but 
%also  on which walls does graviton zero mode localizes.

In global SUSY theory there are many models which have 
exact soliton solutions, such as domain wall junction 
\cite{OINS}--\cite{KakimotoSakai}. 
A good progress has been made to study 
domain wall junction in chiral scalar fields 
coupled to SUGRA \cite{CHT2}, 
although no explicit solution has been 
obtained so far. 
It is an intriguing problem to extend 
our prescription for such models and to give 
exact solutions for the soliton. 
This is a remaining future problem.

\begin{acknowledgments}
We thank Nobuhito Maru and Tsuyoshi Sakata for a 
collaboration from which 
this study has started. 
One of the author (M.E.) would like to thank M.Naganuma 
 for useful discussions. 
The authors thank Nobuyoshi Ohta for pointing out two misprints 
in our original manuscript and a useful discussion. 
This work is supported in part by Grant-in-Aid for 
Scientific Research from the Ministry of Education, 
Culture, Sports, Science and Technology, Japan,  
No.13640269. 
One of the authors (M.E.) 
gratefully acknowledges 
support from the Iwanami Fujukai Foundation.
\end{acknowledgments}

\appendix

\section{Uniqueness of the superpotential}

Here we shall show that our prescription of gravitationally 
deformed superpotential is the only possibility to obtain 
BPS solution for matter scalar in SUGRA theories 
which are identical to that in global SUSY theories.¡¡

The BPS equation 
in SUGRA theories comes from requiring 
the vanishing SUGRA transformations in 
(\ref{SUSY_transf}). 
To make this BPS equation for the matter scalar 
identical to the BPS equation 
in global SUSY theories, the only possibility is 
\begin{equation}
\partial_j P_{\rm gl}=
 {\rm e}^{\frac{\kappa^2}{2}K} D_{j}P_{\rm lc}. 
 \label{eq:SUGRA-SUSY}
\end{equation}
For simplicity, we assume that 
only one scalar field has 
nontrivial field configurations in BPS solution, 
and the kinetic term is minimal: 
\begin{equation*}
K(\phi, \phi^*)=\phi^*\phi. 
\end{equation*}
Moreover we assume a real field configuration for the 
BPS solution. 
Then the condition (\ref{eq:SUGRA-SUSY}) reduces to 
\begin{equation*}
{d P_{\rm gl}(\phi) \over d\phi}=
{d \over d\phi}
\left({\rm e}^{{\kappa^2\phi^2 \over 2}}
P_{\rm lc}(\phi)\right). 
\end{equation*}
We obtain the general solution with an integration constant $a$ 
\begin{equation*}
{\rm e}^{{\kappa^2\phi^2 \over 2}}P_{\rm lc}(\phi)
 =
 P_{\rm gl} (\phi)+a. 
\end{equation*}
The assumed reality of field configuration requires 
that the integration constant $a$ should be real.

%\newpage

\end{document}